\documentclass[conference]{IEEEtran}
\IEEEoverridecommandlockouts
\usepackage{cite}
\usepackage{amsmath,amssymb,amsfonts}
\usepackage{algorithmic}
\usepackage{graphicx}
\usepackage{textcomp}
\usepackage{xcolor}
\usepackage{booktabs}
\usepackage{siunitx}
\usepackage{caption}
\usepackage{textcomp}
\usepackage{atbegshi}
\usepackage{hyperref}

\usepackage{float}
\usepackage{placeins}

\def\BibTeX{{\rm B\kern-.05em{\sc i\kern-.025em b}\kern-.08em
    T\kern-.1667em\lower.7ex\hbox{E}\kern-.125emX}}
\IEEEoverridecommandlockouts

\begin{document}

\title{Improving the Performance of Sequential Recommendation Systems with an Extended Large Language Model}

\author{
\IEEEauthorblockN{Sinnyum Choi}
\IEEEauthorblockA{
\textit{Department of AI Convergence Software} \\
\textit{Dong-Seoul University} \\
Seoul, Korea \\
laststar0203@gmail.com}
\and
\IEEEauthorblockN{Woong Kim\textsuperscript{*}}
\IEEEauthorblockA{
\textit{Department of AI Convergence Software} \\
\textit{Dong-Seoul University} \\
Seoul, Korea \\
woong2241@korea.kr}
\thanks{\textsuperscript{*}Corresponding author: woong2241@korea.kr}
}
\maketitle
\IEEEpubidadjcol
\begin{abstract}
Recently, competition in the field of artificial intelligence (AI) has intensified among major technological companies, resulting in the continuous release of new large-language models (LLMs) that exhibit improved language understanding and context-based reasoning capabilities. It is expected that these advances will enable more efficient personalized recommendations in LLM-based recommendation systems through improved quality of training data and architectural design. However, many studies have not considered these recent developments. In this study, it was proposed to improve LLM-based recommendation systems by replacing Llama2 with Llama3 in the LlamaRec framework. To ensure a fair comparison, random seed values were set and identical input data was provided during preprocessing and training. The experimental results show average performance improvements of 38.65\%, 8.69\%, and 8.19\% for the ML-100K, Beauty, and Games datasets, respectively, thus confirming the practicality of this method. Notably, the significant improvements achieved by model replacement indicate that the recommendation quality can be improved cost-effectively without the need to make structural changes to the system. Based on these results, it is our contention that the proposed approach is a viable solution for improving the performance of current recommendation systems.
\end{abstract}


\section{Introduction}
Advances in artificial intelligence have driven technological innovation across a wide range of industries, and recommendation systems are at the center of this transformation. Conventional recommendation systems focus on analyzing user-behavior data to generate recommendations, primarily through conventional methods such as collaborative filtering or content-based filtering. However, these methods have several limitations, such as the small amount of data and the problem of cold-start for new users and items.

To overcome these limitations, artificial neural networks (ANN) have been introduced, which can integrate various forms of data (text, images, temporal information, etc.) and learn complex nonlinear relationships, greatly improving the accuracy and generalization performance of recommendations. To compensate for the difficulties of conventional recommendation systems in accounting for the passage of time, models from the recurrent neural network (RNN) family have been developed to analyze complex user behavior patterns over time and enable sophisticated sequential recommendations.

Recently, Large Language Models (LLMs) have been actively used in recommendation systems to understand complex text-based contexts that are difficult to handle with conventional recurrent neural network-based models. LLM-based recommendation systems are attracting attention because they embed a vast amount of knowledge through pretraining, enabling them to understand user requests and contexts more deeply and make more advanced personalized recommendations.

LLMs are deep learning models based on Transformer \cite{Vaswani2017}. They are trained on large amounts of text data to understand context and generate text. LLMs are pretrained on large-scale, high-quality datasets collected from various domains, exceptional contextual understanding.

Thanks to this capability, LLM-based recommendation systems are able to effectively understand contextual information, including user queries and item descriptions, and thus capture the complex relationships between users and items \cite{Geng2022}. In addition, these systems can leverage their pretrained knowledge to make meaningful predictions, even with limited data on recommendation tasks \cite{Kang2023}. The powerful reasoning capabilities of LLMs can help improve both the personalization and diversity of recommendations \cite{Wu2024}.

As the core of LLM-based recommendation systems, LLMs have evolved rapidly since the introduction of ChatGPT by OpenAI in 2022 \cite{OpenAI2022}, driven by the intensifying AI competition between major technology companies. Every year, new versions of LLMs are released, featuring expanded model architectures and improvements in the quality of training data, resulting in improved reasoning capabilities. These circumstances have had a positive effect on research on LLM-based recommendation systems.

However, a review of the literature on LLM-based recommender systems shows that they have not kept pace with advances in LLM technology \cite{Wu2024}. We hypothesize that modern LLM models are fundamentally superior to earlier versions of LLMs, and that replacing an existing system with an improved LLM can significantly improve recommendation performance, which would be a cost-effective way to improve recommendation systems.

To test this hypothesis, Llama, an open-source LLM developed by Meta was used. Despite being an open-source model, Llama has shown competitive performance compared to commercial LLM models. Since its initial release in 2023, new versions have been released periodically through continuous research and development. Most recently, Llama3 \cite{Grattafiori2024}, the successor to Llama2 \cite{Touvron2023}, was released.

Key improvements in Llama3 include an approximately 8.7-fold increase in pre-training data to 15.6T tokens and the application of Grouped Query Attention \cite{Ainslie2023} to the model architecture, which improves inference speed and memory efficiency. Benchmark evaluations of Llama3’s performance showed improvements across all metrics. In particular, Llama3 achieved a score of 87.3 on the MMLU benchmark, which is close to GPT-4.

In this study, we hypothesize that by simply replacing the model with a recommendation system based on Llama2 to Llama3, personalized recommendations can be made more sophisticated and faster owing to improvements in the quality of the training data and model architecture.

To analyze this experimentally, LlamaRec was used \cite{Yue2023}, which has shown high performance in previous studies. LlamaRec combines a sequence recommendation model with a fine-tuned version of the LLM model from Llama2-7B to form a two-level recommendation system that improves the recommendation performance. The overall working procedure of LlamaRec was retained but only the LLM was replaced with Llama3 to compare and analyze the changes in recommendation performance.

This study has empirically shown that replacing LLM-based recommendation systems with improved LLMs can lead to meaningful improvements in recommendation performance. It also emphasizes the need to actively incorporate the latest LLMs in both the research and practical phases of recommendation systems as new LLMs are constantly being released. Furthermore, it is suggested that future advances in LLM technology will positively affect the performance of LLM-based recommendation systems.

\section{Related works}

\subsection{Large language model}

In recent years, LLMs have continued to evolve and expand their applicability to different domains. OpenAI introduced GPT-4o \cite{OpenAI2024}, a lightweight version of GPT-4 that maintains the same level of language understanding and generation capabilities, but significantly improves inference speed and efficiency. This makes GPT-4o suitable for real-time applications, further expanding the scope of LLM. 

Similarly, Google DeepMind released Gemini 2.5 Pro \cite{Google2025}, which offers twice the inference speed of its predecessor, Gemini 1.5 Pro, and shows improved performance in benchmark evaluations. Anthropic introduced the Claude 3.5 Sonnet \cite{Anthropic2024}, which outperformed other LLMs in tasks such as text generation and coding. These advances in the commercial LLM ecosystem highlight ongoing competition between the major technology companies, all striving to push the boundaries of LLM technology through proprietary models. This competitive situation is driving innovation in this field.

Open-source LLM ecosystems have evolved significantly in recent years. For example, meta built on the success of Llama2 and introduced Llama3 \cite{Grattafiori2024}. Llama3 leverages high-quality data more effectively and features an optimized architecture, resulting in substantial improvements in both performance and efficiency. The latest version, Llama3.2, extends its capabilities by supporting multimodals, broadening its range of applications. These technological advances in LLMs have provided a critical foundation for their expansion into various fields, including recommendation systems.

\subsection{LLM-based recommendation system}

LLMs have been applied to recommendation systems and have contributed significantly to improving recommendation performance \cite{Wu2024}. RecSysLLM retains the general common sense and reasoning capabilities of conventional LLMs but introduces a method for pretraining with data specific to the recommendation domain and has shown strong performance on zero-shot. Experimental results showed that RecSysLLM outperformed existing models in several recommendation benchmarks, demonstrating its adaptability and effectiveness in diverse recommendation tasks \cite{Chu2023}. TallRec argued that LLM's in-context learning is not optimized for the recommendation task and presented results showing that instruction turning improves recommendation performance and maximizes efficiency by using only a limited amount of data \cite{Bao2023}.

As different methods for applying LLM to recommendation systems have been proposed, experimental studies have been conducted on publicly available learning methods. Few-shot and zero-shot learning methods have shown that the LLM is limited because it requires a large number of parameters to replace existing recommendation models. However, fine-tuning has shown that the LLM can perform well with less data compared to existing recommendation models, and experiments with two models with different parameter sizes have confirmed that parameter size can affect recommendation performance \cite{Kang2023}.

Using instruction tuning as fine-tuning, LlamaRec proposed a new framework with a two-stage approach that combines LLM with a sequence recommendation model and further optimizes the auto-regression operation of LLM to improve the efficiency of inference \cite{Yue2023}. This two-step approach effectively integrates LLMs with sequential modeling and improves recommendation accuracy and inference efficiency.

Based on the progress of these studies, our focus was on the benefits of fine-tuning LLM-based recommendation systems and LlamaRec was selected as the experimental framework.

\begin{figure*}[!htbp]
\centering
\includegraphics[width=\textwidth]{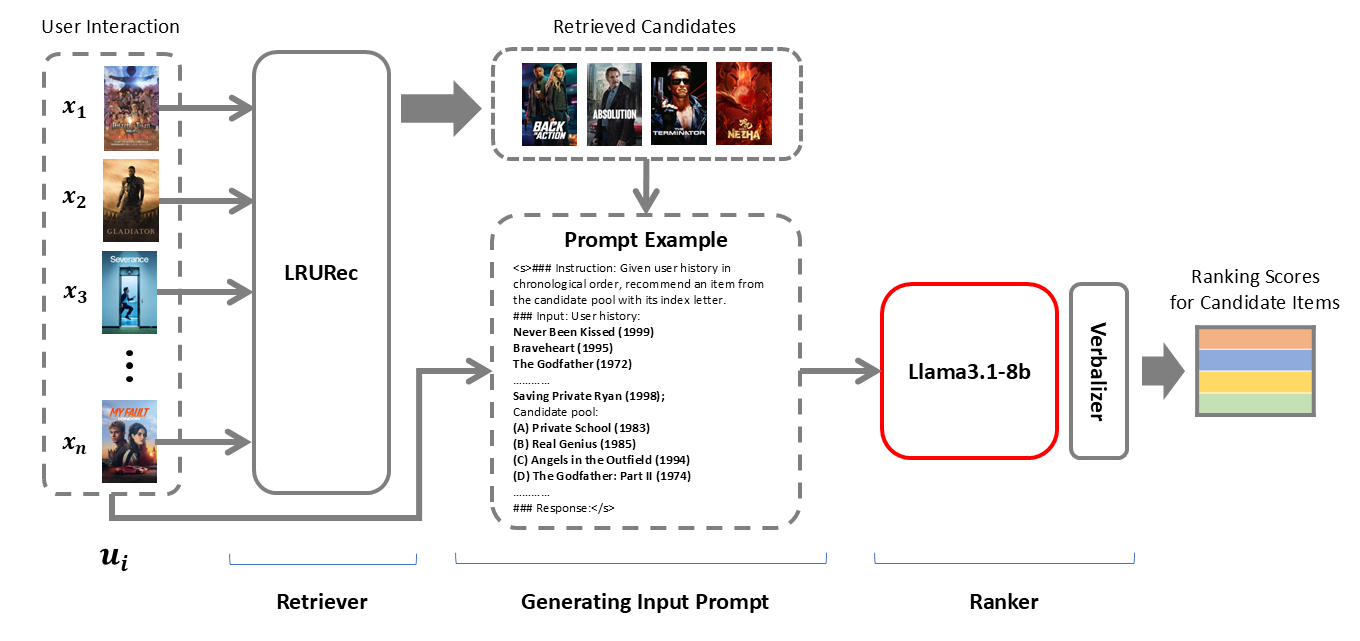}
\caption{LlamaRec processing flowchart. The areas highlighted in red in the figure represent the modified areas}
\label{fig:figure1}
\end{figure*}

\FloatBarrier

\section{Methodology}

\subsection{Based framework}

First, LlamaRec is briefly described, which was selected as the underlying framework for our experiments. LlamaRec uses a two-stage recommendation system commonly used in recommendation systems that work with large datasets \cite{Covington2016, Higley2022}.

LlamaRec consists of two stages: the retriever and the ranker. The retriever is responsible for identifying candidate items. It takes the user's purchase history as input and computes recommendation scores for all items. The user's purchase history can be represented as a vector: \(u_{i} = [x_1, x_2, x_3, \cdots, x_n] \) where \( x_j \) denotes the \(j\)-th item in user \(i\)’s purchase history. The retriever uses this input to generate a recommendation score vector for user \(i\): \( f_{retriever}(u_{i}) \)

Here, \( f_{retriever}(u_{i}) \) represents the recommendation score vector predicted by the retriever for all items for user . Based on these scores, the retriever ranks the items and selects the top-N most relevant ones. These selected items are referred to as the retrieved candidates for user \(i\), denoted as \(c_i\)

The ranker then re-ranks the retrieved candidates. This can be formally represented as: \( f_{ranker}(c_{i}) \).  It uses the LLM for this process and provides input prompts that reflect the metadata of the candidate items. A custom verbalizer converts the logit values of the index characters corresponding to the candidate items into a probability distribution and applies argmax to select the best item.

LlamaRec is shown in Figure~\ref{fig:figure1}. The areas highlighted in red represent modified areas.

\subsection{Proposed framework}

The modifications made to the learning process of LlamaRec for our experiment are described in three steps.

\subsubsection{Preprocessing for the retriever training and retriever training}

In this step, the raw data is preprocessed and the retriever is trained. LRURec was used as the retriever in the framework \cite{Yue2024}. As the retriever performance affects the overall recommendation performance, the hyperparameters were tuned for each dataset separately. The hyperparameters set include weight\_decay and dropout.

\subsubsection{Preprocessing for the ranker training}

\begin{figure}[H]
\centering
\includegraphics[width=.45\textwidth]{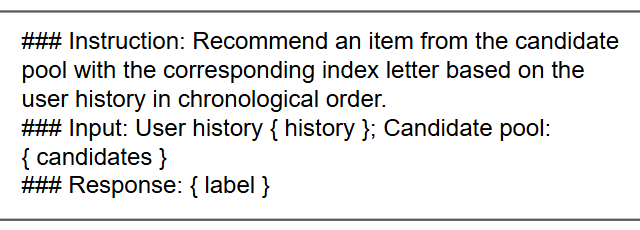}
\caption{Prompt structure}\label{fig:figure2}
\end{figure}

In this step, the data are preprocessed to train the ranker LLM. The prompt structure of the training data, which is used in LlamaRec, is shown in Figure~\ref{fig:figure2}.

Each history and candidate inserts a list consisting of item titles, as shown in Figure~\ref{fig:figure1}, and each candidate also contains an index letter (e.g., A or B) assigned to the item. In the learning process, the candidate is not used with the actual retrieved candidates but with sample data generated by the negative sampling technique, for which random sampling is used. The final generated text prompts need to be tokenized and marked as masked for LLM input. In this process, the tokenizer was changed to a version that is compatible with the new LLM.

\subsubsection{Ranker training}

In this phase, the ranker was trained. The training data generated in the previous step was used for the fine tuning. The main difference is that the existing system uses the lightweight Llama2-7b-hf model, whereas our framework uses a similarly sized Llama3.1-8b model. Although Llama3.2 was the latest version at the time of writing, the difference in text inference performance was not significant. Therefore, Llama3.1-8b was chosen with relatively small parameters for learning efficiency.

In addition, the QLoRA technique was used for LLM learning in the existing framework. Previous research \cite{Huang2024} has shown that applying the same technique to Llama3 can significantly reduce the learning time while maintaining the same performance. Therefore, the QLoRA technique was used in this study.

\section{Experiments}

The objective of this study was to observe the change in performance due to the replacement of the LLM by maintaining the methodology of the existing LlamaRec while changing the LLM only at the ranker stage. This approach allowed us to analyze the effect of LLM replacement on the performance of the recommendation system. The experimental procedure applied in this study is as follows Figure~\ref{fig:figure3}:

\begin{figure}[H]
\centering
\includegraphics[width=.35\textwidth, height=8cm]{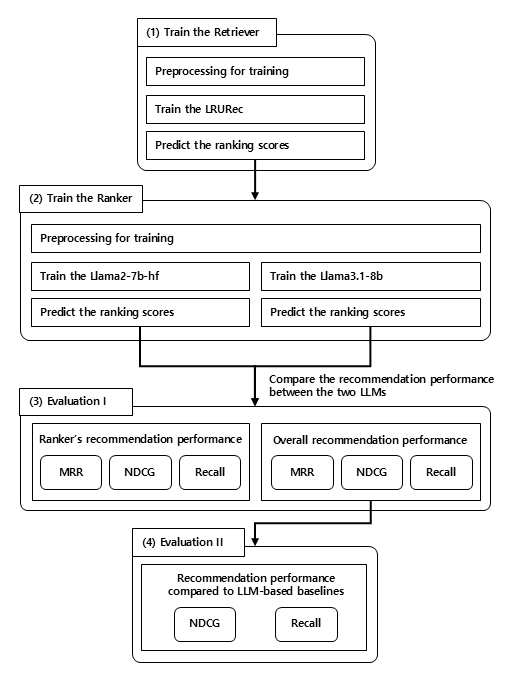}
\caption{Experimental Process}\label{fig:figure3}
\end{figure}

In addition, to ensure the consistency and reproducibility of the experiments, the random seed value was set to 42 throughout the process to ensure that the experimental results are not influenced by external factors and reflect changes that are solely due to the differences in the LLMs. Based on this rigorous experimental design, this study aimed to evaluate the effects of LLM replacement on recommender system performance more accurately and objectively.

\subsection{Dataset}

In this study, the ML-100K, Beauty, and Games datasets were used, which are commonly used to evaluate recommendation system models.

\begin{itemize}
  \item \textbf{ML-100K\cite{Harper2015}:} This dataset consists movie rating data provided by MovieLens and contains 100K user-movie ratings. It contains information such as user ID, movie ID, rating, and timestamp and is often used to evaluate recommendation system algorithms.
  
  \item \textbf{Beauty\cite{He2016,McAuley2015}:} This dataset contains review data for cosmetics and beauty products sold on Amazon. A given review contains the user ID, product ID, rating, the text of the review, etc. 

  \item \textbf{Games\cite{He2016,McAuley2015}:} This dataset contains user-review data for video games and related products on Amazon. A given review contains the user ID, product ID, rating, the text of the review, etc.
\end{itemize}

The three datasets were preprocessed according to previous research methods \cite{Yang2023, Yue2023}. For each dataset, a descending chronological sort by timestamp value was performed, sequences with less than five user-item interactions for training data quality were removed, and items without metadata were also removed. Table~\ref{tbl:table1} lists these statistics.

\begin{table}[htbp]  % h:here, t:top, b:bottom, p:page
\centering
\setlength\tabcolsep{0.9em}
\caption{Dataset statistics after preprocessing}
\label{tbl:table1}
\begin{tabular}{c|ccccc}  % 열 정렬: c=center, l=left, r=right, |는 선
\hline
\textbf{Dataset}  & \textbf{Users} & \textbf{Items} & \textbf{Interact} & \textbf{Length} & \textbf{Density} \\ \hline
\textbf{ML-100K}  & 610     & 3,650   & 100K     & 147.99   & 4e-2  \\ 
\textbf{Beauty}   & 22,332  & 12,086  & 198K     & 8.87     & 7e-4  \\ 
\textbf{Games}    & 15,264  & 7,676   & 148K     & 9.69     & 1e-3  \\ 
\hline
\end{tabular}
\vspace{-1em}
\end{table}

\subsection{Baseline method}

The primary objective of this study was to compare the performance of LlamaRec with that of LlamaRec using Llama3. In addition, methods adopted from previous studies (i.e., LlamaRec) were included as baseline models to further support our evaluation. All included methods have been published in the literature on LLM-based recommendation systems.

\begin{itemize}

  \item \textbf{LlamaRec\cite{Yue2023}:} This is a framework developed for two-stage ranking-based recommendation using LLMs.
  \item \textbf{PALR\cite{Yang2023}:} This framework combines user behavior data with LLMs to make personalized recommendations and consists of three steps: natural-language-based user profile generation, candidate discovery, and LLM ranking.
  \item \textbf{GPT4Rec\cite{Li2023a}:} This framework leverages a GPT-2-based generative model and a BM25 search engine to translate user interests into interpretable search queries, thereby improving the variety and accuracy of personalized recommendations.
  \item \textbf{RecMind\cite{Wang2023}:} This is an LLM-based recommendation agent and autonomous recommendation system that uses Self-Inspiring (SI) planning techniques to combine multi-level reasoning and external knowledge to perform various recommendation tasks.
  \item \textbf{POD\cite{Li2023b}:} This is a Prompt Distillation technique for LLM-based recommendation systems that provides a model-independent approach to optimize recommendation performance and improve learning and inference efficiency by leveraging continuous prompt vectors.
\end{itemize}

\subsection{Evaluation}

In this study, a dataset created with the leave-one-out strategy was used for the evaluation. Mean Reciprocal Rank (MRR@\(k\)), Normalized Discounted Cumulative Gain (NDCG@\(k\)), and Recall@\(k\) were used, where \(k\) was set to two values: 5 and 10. NDCG@10 was selected as the metric for the best validation score when training the ranker.

\subsubsection{Ranker's recommendation performance}

t uses a validation dataset to compare the change in performance with the LLM replacement. Let \(U\) denote the entire set of users. The subset \( R \subseteq U \) represents users for whom the retriever’s top-k predictions contain the correct label, and \( \bar{R} \) represents those for whom it does not. These subsets satisfy the following relation in:

\begin{equation}
U = R \cup \bar{R}, \quad R \cap \bar{R} = \emptyset
\end{equation}

To evaluate the ranker’s performance, we consider the recommendation scores predicted by the ranker for each candidate set \(c_k\) where \( k \in R \) aggregated as:

\begin{equation}
S_{R} = \left[ f_{ranker}(c_{k}) \right]_{k \in R} 
\end{equation}

The evaluation metrics are then defined as a vector of standard ranking measures in: 

\begin{equation}
M(x) = \left[ 
\begin{array}{c}
\mathrm{MRR}@5(x) \\
\mathrm{NDCG}@5(x) \\
\mathrm{Recall}@5(x) \\
\mathrm{MRR}@10(x) \\
\mathrm{NDCG}@10(x) \\
\mathrm{Recall}@10(x)
\end{array} 
\right]
\end{equation}

The Ranker’s performance is computed as: 

\begin{equation}
P_{ranker} = M(S_{R}) 
\end{equation}

\subsubsection{Overall recommendation performance}

To evaluate the overall system performance, we also take into account the users in \( \bar{R} \)
for whom the retriever failed to include the correct label in the top-k predictions. For each user \( k \in \bar{R} \) the recommendation score metics predicted by the retriever is defined as: 

\begin{equation}
S_{\bar{R}} = \left[ f_{retriever}(u_{k}) \right]_{k \in \bar{R}}
\end{equation}

The overall recommendation performance across the full user set \( U = R \cup \bar{R} \) is then computed using a weighted average as: 

\begin{equation}
P_{overall} = \frac{ M(S_{R}) \cdot \left| R \right| + M(S_{\bar{R}}) \cdot \left| \bar{R} \right| }{ \left| U \right| }
\end{equation}

These equations (1) to (6) provide a formal framework for evaluating both the ranker's and the overall system's recommendation performance. This enables comprehensive evaluation by combining retriever and ranker performance.

\subsection{Implementation}

In this study, the model was implemented using the PyTorch framework. Llama3 and tokenizer were used by loading “meta-llama/Llama3.1-8b” registered in the Hugging Face with the Python Transformers module.

To train the retriever (i.e. LRURec), the decay was set to 0 and dropout to 0.5 for the ML-100K dataset, whereas the decay was set to 0.01 and the dropout to 0.5, for the other datasets. The other hyperparameters were the same as those used in a previous study.

To train the ranker (i.e. Llama3.1-8b), a maximum of 20 historical items were used and the 20 best candidates were selected from the retriever results. When applying QLoRA, we set the number of LoRA dimensions to 8, the value of \(a\) to 32, and the dropout to 0.05. We set the LoRA learning rate to 1e-4, and the modules to be tuned are the \(Q\) and \(V\) projection matrices. The batch size was set to 16 for the ML-100K dataset and 12 for the other datasets, which is consistent with the results of previous studies. The epoch was set to 1 and validated every 100 iterations. The model with the best validation performance is saved for the evaluation of the test set.

\section{Result}

\subsection{Recommendation performance}

The recommendation performance was evaluated before and after LLM replacement. The results of the evaluation of recommendation performance for the ranker are presented in Table~\ref{tbl:table2}, and the overall evaluation of recommendation performance is presented in Table~\ref{tbl:table3}. In these two tables, each row represents an evaluation metric and each column represents a recommendation method (and dataset). M, N, and R are abbreviations for MRR, NDCG, and RECALL, respectively, and the best results are in bold font.

\begin{table*}[ht]
\centering
\setlength\tabcolsep{1.7em}
\caption{Ranker's recommendation performance}
\label{tbl:table2}
\resizebox{\textwidth}{!}{
\begin{tabular}{c|cccccc}
\hline
\textbf{Metric} & \multicolumn{2}{c}{\textbf{ML-100K}} & \multicolumn{2}{c}{\textbf{Beauty}} & \multicolumn{2}{c}{\textbf{Games}} \\
 & LlamaRec & \textbf{Our} & LlamaRec & \textbf{Our} & LlamaRec & \textbf{Ours} \\
\hline
\textbf{M@5}   & 0.1728 & \textbf{0.2961} & 0.2972 & \textbf{0.3254} & 0.2805 & \textbf{0.3130} \\
\textbf{N@5}   & 0.2317 & \textbf{0.3462} & 0.3433 & \textbf{0.3795} & 0.3379 & \textbf{0.3726} \\
\textbf{R@5}   & 0.4126 & \textbf{0.5} & 0.4835 & \textbf{0.5446} & 0.5136 & \textbf{0.5542} \\
\textbf{M@10}  & 0.2083 & \textbf{0.3235} & 0.3294 & \textbf{0.3545} & 0.3129 & \textbf{0.3428} \\
\textbf{N@10}  & 0.3184 & \textbf{0.4142} & 0.4217 & \textbf{0.4503} & 0.4169 & \textbf{0.4455} \\
\textbf{R@10}  & 0.6825 & \textbf{0.7142} & 0.7273 & \textbf{0.7641} & 0.7588 & \textbf{0.7810} \\
\hline
\end{tabular}
} 
\end{table*}

As shown in Table~\ref{tbl:table2}, when the original ranker model Llama2-7b was replaced by Llama3.1-8b the recommendation performance improved across all metrics. Compared to the existing method, LlamaRec, average performance improvements of 38.65\%, 8.69\%, and 8.19\% were achieved on the ML-100K, Beauty, and Games datasets, respectively. However, a closer look at the detailed evaluation results shows that the Recall@5 metric has a relatively low performance improvement, which is important considering that NDCG@10 was used as the evaluation ranking metric in the training process. When evaluating with NDCG@10, the performance improvements were 30.09\%, 6.77\%, and 6.87\% on the ML-100K, Beauty, and Games datasets, respectively.

\begin{table*}[ht]
\centering
\setlength\tabcolsep{1.7em}
\caption{Overall recommendation performance}
\label{tbl:table3}
\resizebox{\textwidth}{!}{
\begin{tabular}{c|cccccc}
\hline
\textbf{Metric} & \multicolumn{2}{c}{\textbf{ML-100K}} & \multicolumn{2}{c}{\textbf{Beauty}} & \multicolumn{2}{c}{\textbf{Games}} \\
 & LlamaRec & \textbf{Our} & LlamaRec & \textbf{Our} & LlamaRec & \textbf{Ours} \\
\hline
\textbf{M@5}   & 0.0358 & \textbf{0.0611} & 0.0368 & \textbf{0.0403} & 0.0587 & \textbf{0.0655} \\
\textbf{N@5}   & 0.0479 & \textbf{0.0715} & 0.0425 & \textbf{0.0469} & 0.0707 & \textbf{0.0780} \\
\textbf{R@5}   & 0.0852 & \textbf{0.1032} & 0.0598 & \textbf{0.0674} & 0.1075 & \textbf{0.1160} \\
\textbf{M@10}  & 0.0431 & \textbf{0.0668} & 0.0407 & \textbf{0.0439} & 0.0655 & \textbf{0.0718} \\
\textbf{N@10}  & 0.0658 & \textbf{0.0855} & 0.0522 & \textbf{0.0557} & 0.0873 & \textbf{0.0933} \\
\textbf{R@10}  & 0.1409 & \textbf{0.1475} & 0.0900 & \textbf{0.0946} & 0.1589 & \textbf{0.1635} \\
\hline
\end{tabular}
}
\vspace{-1em}
\end{table*}

As shown in Table~\ref{tbl:table3}, the overall performance evaluation using the entire dataset shows positive results, reflecting the performance improvement of the ranker. Compared to LlamaRec, average performance improvements of 38.43\%, 8.69\%, and 8.19\% were achieved on the ML-100K, Beauty, and Games datasets, respectively.
In summary, this means that the LLMs that were replaced in our experimental framework more accurately reflect user preferences and contribute to improving the quality of recommendations.

\subsection{Recommendation performance compared to LLM-based baselines}

To evaluate the usability of our framework from an objective perspective, it was compared with other LLM-based recommendation systems. It was compared with Beauty, which has been frequently used as a benchmark in several studies on LLM-based sequential recommendation systems. When the implementation code was not provided, the results were compared with those from previous studies. The LLM-based recommendation methods used in the comparison included PALR, GPT4Rec, RecMind, and POD \cite{Li2023a, Li2023b, Wang2023, Yang2023}.

The results of the comparison are summarized in Table~\ref{tbl:table4}, which is reported in a similar way to Table~\ref{tbl:table3}, with the second-best result underlined.

\begin{table}[ht]
\vspace{0.2em}
\centering
\setlength\tabcolsep{0.9em}
\caption{Recommendation performance compared to LLM-based baselines}
\label{tbl:table4}
\begin{tabular}{c|cccccc}
\hline
\textbf{Metric} & \multicolumn{5}{c}{\textbf{Beauty}} \\
& \textbf{PALR} & \textbf{GPT4Rec} & \textbf{RecMind} & \textbf{POD} & \textbf{Ours} \\
\hline
\textbf{N@5}   & N/A   & N/A    & 0.0289 & \underline{0.0395} & \textbf{0.0469} \\
\textbf{R@5}   & N/A   & \underline{0.0653} & 0.0415 & 0.0537 & \textbf{0.0674} \\
\textbf{N@10}  & \underline{0.0446} & N/A   & 0.0375 & 0.0443 & \textbf{0.0557} \\
\textbf{R@10}  & 0.0721 & \underline{0.0810} & 0.0574 & 0.0688 & \textbf{0.0946} \\
\hline
\end{tabular}
\vspace{-0.5em}
\end{table}

The results show that most of the underlined metrics are the second-best-performing metrics in the literature before LlamaRec, indicating that our framework is overwhelmingly dominant. Notably, PALR's NDCG@10 showed the greatest improvement, an increase of 24.88\%. This significant improvement emphasizes the effectiveness of the proposed approach in terms of ranking performance. While LlamaRec had a worse Recall@5 result than GPT4Rec in the previous study, our framework shows an improvement of 3.22\% over it.

\section{Conclusions}

In this study, the existing Llama2 was upgraded to Llama3 in LlamaRec, an LLM-based recommendation system, and the effect of the improved LLM on recommendation performance was analyzed. The experimental results show that our framework has significant improvements in ranker performance and overall recommendation performance compared to the existing framework. Furthermore, comparisons were made with other LLM-based recommendation systems and an overwhelming performance improvement was demonstrated.

This confirms that upgrading the LLM can effectively improve the recommendation performance, suggesting that replacing the LLM can play an important role in improving the performance of LLM-based recommendation systems. The experimental results of this study show that upgrading LLMs can substantially improve the recommendation quality, implying that similar performance improvements can be expected when applied to different LLM-based recommendation systems in the future. This paper is also significant because it contributes to previous research by improving the performance of rankers in LlamaRec and the overall quality of recommendation.

We anticipate that in future research, we will conduct the same experiments with different LLM-based recommendation systems based on the results of this study to investigate the generalizability of the results and conduct an in-depth analysis of various aspects, not just recommendation performance.

%Bibliography
\bibliographystyle{unsrt}  
\bibliography{main}  

\end{document}